\documentclass[final]{iopart}
\usepackage{iopams}
\usepackage{graphicx}
\usepackage{bm}
\usepackage{color}
\usepackage{graphicx,epsfig}
\usepackage{amsfonts,amssymb}

\newtheorem{theorem}{Theorem}[section]
\newtheorem{corollary}{Corollary}[section]

\newcommand{\fs}{f_{\mathrm{s}}}

\newcommand{\tc}{T_{\mathrm{c}}}

\newcommand{\spek}{\mathrm{spec}}

\renewcommand{\Re}{\mathrm{Re}}

\newcommand{\regsol}{\varphi} \newcommand{\regsolo}{\mathring{\varphi}}
\newcommand{\kernK}{K}
\newcommand{\kernP}{P}
\newcommand{\zm}{\mathsf{z}} \newcommand{\zL}{\frak{z}}
\newcommand{\sk}{k} \newcommand{\sko}{\mathring{\sk}}
\newcommand{\normreg}{\varkappa}
\newcommand{\eqref}[1]{(\ref{#1})}
\newcommand{\normrego}{\mathring{\varkappa}}

\definecolor{dblue}{rgb}{0.1,0.1,0.44}
\definecolor{dgreen}{rgb}{0.2 ,0.54, 0.2}

\newcommand{\C}[1]{{}}
\newcommand{\tfrac}[2]{\textstyle{\frac{#1}{#2}}}

\newcommand{\aseq}{\simeq}
\newcommand{\asprop}{\sim}
\newcommand{\etwa}{\approx}

\eqnobysec
\begin{document}

\paper[Inverse scattering theory and trace formulae\ldots]
{Inverse scattering theory and trace formulae for one-dimensional Schr\"odinger problems with singular potentials}
\author{S.~B.\ Rutkevich and H.~W.  Diehl}
\address{%
  Fakult\"at f\"ur Physik, Universit{\"a}t Duisburg-Essen, D-47048 Duisburg,
  Germany}
\begin{abstract}
Inverse scattering theory is extended to one-dimensional Schr\"odinger problems with near-boundary singularities of the form $v(z\to 0)\aseq  -z^{-2}/4+v_{-1}z^{-1}$. Trace formulae relating the boundary value $v_0$ of the nonsingular part of the potential to spectral data are derived. Their potential is illustrated by applying them to a number of Schr\"odinger problems with singular potentials.
\end{abstract}

\section{Introduction}
Inverse scattering theory, developed more than 60 years ago by Gel'fand, Krein, Levitan, Marchenko, and others \cite{GL51,GL53,Kre53,Run97,Cha97,CS89}, enables one to reconstruct potentials of Sturm-Liouville or Schr\"odinger problems from scattering data provided they satisfy certain mathematical conditions \footnote{For background on inverse scattering theory and an extended list of references, see \cite{CS89}.}. In the case of Sturm-Liouville problems on the half-line, the potential $v(z)$ is typically  required to be continuous, to decay sufficiently fast at infinity so that it is absolutely integrable, and to remain finite on approaching the boundary $z=0$.  In a recent paper \cite{RD15} (hereafter referred to as I), we investigated the exact  solution of the $O(n)$ $\phi^4$ theory on a three-dimensional  film bounded in one direction by a pair of planar free surfaces in the many-component limit $n\to\infty$. As has been recognised a long time ago in studies of the semi-infinite case where one surface plane is located at $z=0$ while the other is at $z=\infty$, the exact ${n=\infty}$ solution of this model at the bulk critical temperature $\tc$ leads to a one-dimensional Schr\"odinger equation with a singular potential $v(z)=-z^{-2}/4$ \cite{BM77a,BM77c}. For {temperatures}  above and  below $\tc$, one must deal with (rescaled) potentials that minimise a free-energy functional. The solutions of the corresponding equations for these potentials involve the eigenfunctions and eigenvalues of the Sturm-Liouville problem with the same potential. In other words, the potentials are self-consistent solutions of these equations. Following a standard practice in the physics community, we will occasionally indicate this fact by referring to them as ``self-consistent'' potentials.

Unfortunately, the self-consistent potential that gives the exact solution of  $O(n)$ $\phi^4$ model in the limit $n\to\infty$ is known only in numerical rather than in closed analytical form, barring the semi-infinite case at $\tc$ \cite{DR14,DGHHRS12,DGHHRS14,DGHHRS15}. From consistency requirements with the behaviour at $\tc$ and results obtained via boundary-operator expansions \cite{DR14,DGHHRS12,DD81a,Die86a,Die97} {one} can conclude that  they must vary as
\begin{equation}\label{eq:singpot}
v(z\to 0)=-\frac{1}{4z^2}+\frac{v_{-1}}{z}+v_0+\Or(z)
\end{equation}
on approaching the ${z=0}$ surface plane and have a similar singular near-boundary behaviour at the second plane. 

The aim of this paper is to explore how standard tools of  inverse scattering theory such as the Povzner-Levitan representation and the Gel'fand-Levitan equation  \cite{CS89} can be extended to one-dimensional Schr\"odinger problems on the half-line and finite intervals with potentials that exhibit the singular behaviour~\eqref{eq:singpot} at the boundary. In addition, we will derive  trace formulae for  Sturm-Liouville problems with such potentials
whose combination with inverse scattering tools will enable us to express the potential coefficient $v_0$ through scattering data.  In I, we applied some of the results obtained here to the analysis of the exact ${n\to\infty}$ solution of the above-mentioned $O(n)$ $\phi^4$ model, using them to determine a number of quantities exactly. Specifically, we there used trace formulae to be derived below and their corollaries to determine the exact value of the expansion coefficient $v_0$ in equation~\eqref{eq:singpot}. Although trace formulae of a similar kind have been discussed in the literature both for nonsingular potentials as well as certain types of singular potentials, no trace formula for the case of potentials with the particular singular near-boundary behaviour~\eqref{eq:singpot} was available.

To put things in perspective and clarify the basis of knowledge from which we can start, it will be helpful to briefly recall the relevant literature. Singular potential terms of the form $\mu z^{-2}$ with $\mu=l(l+1)$, $l=0,1,\ldots$ naturally arise through the centrifugal contribution to the effective potential of the reduced radial Schr\"odinger equations of radially symmetric three-dimensional Schr\"odinger problems.  The corresponding  inverse scattering problems with nonnegative integer values of the angular momentum quantum number $l$ have been extensively studied for more than 6 decades, see \cite{CS89} and its references.  More recently, Sturm-Liouville problems pertaining to radial Schr\"odinger equations with non-integer values of  $l$ have also attracted considerable interest \cite{Zho94,FY05,Kostenko2010,Kostenko2011,Kostenko2012}. 
Freiling and Yurko  \cite{FY05} investigated  inverse spectral problem for the case of non-integer $l>-1/2$. They described how the potential can be recovered from the spectral data, and proved  a uniqueness theorem. Potentials with  boundary singularities of the form specified in equation~\eqref{eq:singpot} (corresponding to the radial Schr\"odinger equation with $l=-1/2$), have previously been considered by two other groups of authors. Zhornitskaya and  Serov \cite{Zho94} established a  uniqueness theorem for the corresponding inverse spectral problem with vanishing potential coefficient $v_{-1}=0$. Kostenko, Sakhnovich and Teschl \cite{Kostenko2010} succeeded to do this for the more general case $v_{-1}\ne0$. Unfortunately, neither {of} these two papers clarified the question of how the considered  singular potentials can be recovered from the spectral data.

Even if the potential does not become singular at the boundaries, the process of recovering it from scattering or spectral data is not easy because  it usually requires the solution of integral equations \cite{CS89}. If the potential 
 $v(z)$ is analytic near the origin $z=0$, one can try to solve the simpler problem of determining one or several leading contributions of the potential's Taylor expansion about $z=0$ from the scattering or spectral data.
 
 The first ``trace formula'' of this kind  was obtained by Gel'fand and Levitan \cite{GL53}, who expressed the boundary value of a regular potential of the finite-interval Sturm-Liouville problem in terms of  the corresponding eigenvalues. This seminal paper  has triggered  a  lot of  interest in such trace formulae owing to their rich applications in different fields of mathematics and physics, among them integrable nonlinear Hamiltonian systems.  The Gel'fand-Levitan trace formula \cite{GL53}  was generalised in a variety of ways by Dikii \cite{Dikii58}, Newton \cite{Newton55}, Faddeev and Buslaev \cite{Fad57,BusFad60}, and others. 
An extensive list of references can be found in \cite{CS89,Rybkin99,Rybkin2001,Ges95}. The trace formula that comes closest to the one we need and shall derive below is the following one due to Newton \cite{Newton55} and Faddeev \cite{Fad57}{ ,} which relates the expansion coefficient $v_0$ of the Laurent series
 \begin{equation}\label{eq:lsingpot}
v_l(z\to 0)=\frac{l(l+1)}{z^2}+v_0+\Or(z)
\end{equation}
of the effective potential $v_l$  of the radial Schr\"odinger equation on the half-line $0<z<\infty$ for angular momentum $l$ to the scattering phase $\eta_l(k)$ and the discrete energy levels $\varepsilon_\nu$. It reads (see equation~(VII.4.2) of reference~\cite{CS89})
\begin{equation}\label{eq:vl0}
v_0=-\frac{8}{(2l+1)\pi}\int_0^\infty d\sk\,\sk \frac{d}{d \sk}\left[\sk\, \eta_l(\sk)\right]+\frac{4}{(2l+1)}\sum_\nu \varepsilon_\nu.
\end{equation}
However, { there are two reasons  why this formula, equation~ \eqref{eq:vl0}, cannot be applied to the case of a  potential 
with the boundary singularity \eqref{eq:singpot}: The first is that the right-hand side of  \eqref{eq:vl0} does not exist for $l=-1/2$; the second is that the 
 first-order pole term $v_{-1}/z$ of the potential  \eqref{eq:singpot} is absent in \eqref{eq:lsingpot}.}

The remainder of this paper is organised as follows. In the next section we begin by defining more precisely the type of Schr\"odinger problems we will be concerned with. Both Schr\"odinger problems on the half-line $(0,\infty)$ and on a finite interval $(0,N)$ will be considered. We then recall some necessary background of inverse scattering theory for nonsingular potentials, discuss {its } generalisation to potentials with near-boundary singularities of the form specified in equation~\eqref{eq:singpot}, and explain the modifications of the required equations of inverse scattering theory implied by these singularities. In Sec.~\ref{sec:traceform}, we turn to the derivation of trace formulae for Schr\"odinger problems with singular potentials on finite intervals $[0,N]$ and the half-line $[0,\infty)$. Both the trace formula and its corollary we exploited in I are proven. In Sec.~\ref{sec:apptrf} we demonstrate the potential of these trace formulae by applying   them to a number of illustrative examples. Section~\ref{sec:conclrem} contains concluding remarks.

\section{Schr\"odinger problems for potentials with singularities at boundaries}\label{sec:SPsingpot}

We will be concerned with the differential equation (Schr\"odinger equation)
\begin{equation}\label{eq:SE}
\mathcal{H}_v\psi(z)=\varepsilon\,\psi(z)
\end{equation}
where $\mathcal{H}_v$ denotes the operator 
\begin{equation}\label{eq:Hv}
\mathcal{H}_v=-\partial_z^2+v(z).
\end{equation}
Here $z$  ranges either over the half-line $(0,\infty)$ or the finite interval $(0,N)$. Note that both $z$ and $N$ are dimensionless variables: they correspond to the distance from $z=0$ and the length of the interval measured in units of an appropriate reference length scale, respectively.\footnote{%
In I, a number of distinct reference lengths scales were encountered, namely, the lattice constant $a$ of the discrete model which we used as a starting point there, the correlation and Josephson coherence lengths $1/|m|$ (depending on whether {the temperature variable $m$ was $\gtrless 0$)}, and the film thickness $L=Na$. In applying the results of {the present} paper to I, the proper identification of the variables  $z$ and $N$ must be made. For example, depending on whether $|m|$ or $L$ is scaled to unity, $z$ corresponds to the variables denoted $\zm$ and $\zL$, respectively, in I.}

In the half-line case we assume the potential to have a Laurent series about ${z=0}$ of the form
\begin{equation}\label{eq:potform}
v(z)=v^{(\mathrm{sg})}(z)+u(z)
\end{equation}
with a singular part
\begin{equation}\label{eq:vsg}
v^{(\mathrm{sg})}(z)=\frac{-1}{4z^2}+v_{-1}\frac{1}{z}
\end{equation}
and a nonsingular contribution (Taylor series part) with the series expansion %
\begin{equation}\label{eq:vTexp}
u(z)=\sum_{j=0}^\infty v_jz^j.
\end{equation}

For the sake of simplicity, we take $u(z)$ to be analytic for all $z\ge 0$. Although the self-consistent potentials that yield the exact large-$n$ solution studied in I are expected to involve non-analytic contributions of order $z^3 \ln z$, this is not a severe restriction and could be relaxed because we shall essentially rely only  on the existence of $u(0)=v_0$ and its first derivative $u^{\prime}(0)=v_1$. We also assume that $v(\infty)=0$ and that $v(z)$ vanishes sufficiently fast for $z\to\infty$ so that 
\begin{equation}
\int_{z_0}^\infty\rmd{z}\,|v(z)|<\infty \mbox{ for all }z_0>0.
\end{equation}

We shall also consider the analogous problems on the finite interval $I=[0,N]$. When dealing with this finite-interval case, we shall assume the potential to have the symmetry property
\begin{equation}\label{eq:vsymmetry}
v(z)={v(N-z)}
\end{equation}
and to be decomposable as in equation~\eqref{eq:potform}. Note that because of the symmetry~\eqref{eq:vsymmetry}, $u(z)$ contains singular boundary terms of the form $v^{(\mathrm{sg})}(N-z)$ unless it is explicitly stated that no singular part is included in  the potential. Depending on the choices of $I=[0,\infty)$ and $I=[0,N]$ and the presence or absence of a singular potential term, we will refer to these cases as half-line or finite-interval problem with singular or non-singular potentials, respectively. To fully define these problems, we must also specify the boundary conditions. 

Consider first the half-line case with non-singular potential. It will be appropriate and sufficient for our purposes to choose Dirichlet boundary conditions $\psi(0)=0$ and assume the absence of bound states.  Extensions to Neumann and Robin boundary conditions are straightforward and are described, for example, in {\cite{Run97} and \cite{CS89}}. Likewise, generalisations to cases in which bound states appear can be found in these review articles and their references.

Let $\varepsilon=\sk^2$ and $\regsol(z,\sk)$ and $\regsolo(z,\sk)$ with $\varepsilon=\sk^2$ denote solutions of this Dirichlet problem on the half-line with two non-singular potentials $v(z)\equiv u(z)$ and $\mathring{v}(z)\equiv \mathring{u}(z)$, respectively, which are normalised such that
\begin{equation}\label{eq:normregsol}
\partial_{z}\regsol(z,\sk)|_{z=0}=1,\quad \partial_{z}\regsolo(z,\sk)|_{z=0}=1.
\end{equation}
These regular solutions are related by a unitary transformation, which entails the so-called  Povzner-Levitan (PL) representation \cite{Run97,Cha97,CS89}
\begin{equation}\label{eq:PLrep}
\regsol(z,\sk)=\regsolo(z,\sk)+\int_0^{z}\rmd{{z}'}\,\kernP(z,{z}')\,\regsolo({z}',\sk).
\end{equation}
Here the PL kernel $\kernP$ depends on the potentials $v$ and $\mathring{v}$, but not on the spectral parameter $\varepsilon$. From rigorous mathematical work (see \cite{CS89}), it is known that the kernel satisfies the conditions \footnote{The condition~\eqref{eq:Pcond} holds when either Dirichlet or Neumann boundary conditions are applied. It gets modified in the case of Robin boundary conditions \cite{Run97}.}
\begin{eqnarray}\label{eq:Pcond}
\kernP(z,z)&=&\frac{1}{2}\int_0^{z}\rmd{s}\big[v(s)-\mathring{v}(s)\big],\\
\kernP(z,0)&=&0,\nonumber
\end{eqnarray}
and the partial differential equation
\begin{equation}\label{eq:diffeqPLkernel}
(\partial_{z}^2-\partial_s^2)\kernP(z,s)=\big[v(z)-\mathring{v}(s)\big]\kernP(z,s),\;\;0\le s\le z.
\end{equation}

A case of particular interest is the choice $\mathring{v}=0$. In this case  the regular solution becomes
\begin{equation}
\mathring{\varphi}(z,\sk)=\frac{\sin(\sk z)}{\sk}.
\end{equation}

As is known from the classical work of Gel'fand and Levitan \cite{GL51} (see also  \cite{CS89}), one can reconstruct from the scattering data the function 
\begin{equation}
 \kernK( z,{ z}')=\int\regsolo( z,\sk)\,\regsolo({ z}',\sk)[\rmd{\rho}(\sk^2)-\rmd\mathring{\rho}(\sk^2)],
\end{equation}
as
\begin{equation}\label{eq:KkernelJost}
\kernK( z,{ z}')=2\int_0^\infty\frac{\rmd{\sk}\,\sk^2}{\pi}\,\regsolo( z,\sk)\,\regsolo({ z}',\sk)\left[\frac{1}{|F(\sk)|^2}-1\right],
\end{equation}
where $F(k)$ is the Jost function, while  $\rmd\rho(\varepsilon)/\rmd\varepsilon$ and $\rmd\mathring{\rho}(\varepsilon)/\rmd\varepsilon$ denote the densities of eigenvalues $\varepsilon=\sk^2$ of the Hamiltonians $\mathcal{H}_v$ and $\mathcal{H}_{\mathring{v}}$, respectively.
The function $\kernK( z, z^\prime)$ can then be used as input for a Fredholm integral equation,  the Gel'fand-Levitan (GL) equation
\begin{equation}\label{eq:GLeq}
\kernK( z,{ z}')+\kernP( z,{ z}')+\int_0^{ z}\rmd{{ z}''}\,\kernP( z,{ z}'')\,\kernK({ z}',{ z}'')=0,
\end{equation}
from whose solution for the PL kernel $\kernP( z,{ z}')$ the potential $v( z)$ can be recovered via 
\begin{equation}\label{eq:ufromP}
v( z)=\mathring{v}( z)+2\frac{\rmd}{\rmd{ z}}\kernP( z, z).
\end{equation}

The above equations carry over to the Dirichlet problem on the  interval $[0,N]$ with non-singular potential, with obvious adjustments. The spectrum becomes discrete{. L}et  $\varepsilon_\nu=\sk^2_\nu$ and $\mathring{\varepsilon}_\nu=\sko_\nu^2$ denote the eigenvalues one obtains for the potentials $v( z)$ and $\mathring{v}( z)$, respectively, and  $\regsol( z,\sk_\nu)$ and $\regsolo( z,\sko_\nu)$ be the associated regular solutions.  Instead of equation~\eqref{eq:KkernelJost}, we then have
\begin{equation}\label{eq:kernKfi}
\kernK( z, z^\prime)=\sum_{\nu=1}^\infty\bigg[\frac{\regsolo( z,\sk_\nu)\,\regsolo( z^\prime,\sk_\nu)}{\normreg_\nu}
-\frac{\regsolo( z,\sko_\nu)\regsolo( z^\prime,\sko_\nu)}{\normrego_\nu}\bigg],
\end{equation}
where $\normreg_\nu$ and $\normrego_\nu$ are the squares of the $L_2([0,N])$ norms of $\regsol( z,\sk_\nu)$ and $\regsolo( z,\sko_\nu)$, e.g.,
\begin{equation}\label{eq:alphanudef}
\normreg_\nu=\int_0^N\rmd{ z}\,|\regsol( z,\sk_\nu)|^2.
\end{equation}
Both the GL equation~\eqref{eq:GLeq} and equation~\eqref{eq:ufromP} remain valid.

We next turn to the cases with singular potentials. We wish to apply the above equations{~\eqref{eq:Pcond}, \eqref{eq:diffeqPLkernel} and \eqref{eq:KkernelJost}--\eqref{eq:ufromP}}  to problems with potentials $v( z)$ of the form specified by equations~\eqref{eq:potform}--\eqref{eq:vTexp} and their analogs $\mathring{v}( z)=v^{(\mathrm{sg})}( z)$ with a vanishing nonsingular part. Whether these equations carry over to potentials with singular parts is a priori not clear.  The original (non-rigorous) derivation of the GL equation in \cite{GL51} at first glance may seem to be generalisable in a straightforward fashion to cases with singular potentials. However, on second { thought} one realises that subtle issues such as the choice of boundary conditions and the interrelated ones of the existence of self-adjoint extensions of the Hamiltonian $\mathcal{H}_v$~\eqref{eq:Hv}  and of the completeness of the set of eigenfunctions arise  for potentials $v(z)$ involving singularities of the form~\eqref{eq:vsg}.

These issues have been investigated in some detail for the critical-point Hamiltonian
\begin{equation}
\mathcal{H}_{\mathrm{c}}\equiv -\partial_{ z}^2-\frac{1}{4 z^2}
\end{equation}
over the interval $[0,1]$ in a recent paper \cite{KLP06}. But this work does not fully cover the cases we are concerned with. First of all, no additional singular term $v_{-1}/ z$ was included. Second, no contribution that becomes singular at the plane $ z=1$ was included. Nevertheless, the paper provides helpful guidance. Let us briefly summarise some of its results relating to our investigations. 

The authors (KLP) of \cite{KLP06} show that the boundary conditions which square-integrable functions $\psi$ belonging to the maximal domain $\mathcal{D}_{\mathrm{max}}(\mathcal{H}_{\mathrm{c}})\equiv\{\psi\in L^2([0,1])\mid\mathcal{H}_{\mathrm{c}}\psi\in L^2([0,1])$ must satisfy at ${ z=1}$ can be parametrised through an angle $\beta_2\in [0,\pi)$ such that they can be written as
\begin{equation}\label{eq:beta2bc}
\psi'(1)\cos\beta_2+\psi(1)\sin\beta_2=0.
\end{equation}
The choices $\beta_2=\pi/2$ and $\beta_2=0$ correspond to Dirichlet and Neumann boundary conditions, respectively.

Owing to the singular behaviour of $\mathcal{H}_{\mathrm{c}}$ at $ z=0$, the boundary conditions {at this plane} must be defined via a limiting procedure. KLP also show that any function $\in \mathcal{D}_{\mathrm{max}}(\mathcal{H}_{\mathrm{c}})$  can be decomposed as
\begin{equation}\label{eq:c12bc}
\psi( z)=[c_1(\psi)+c_2(\psi)\ln z]\sqrt{ z}+\tilde{\psi}( z),
\end{equation}
where $c_j(\psi)$, $j=1,2$, are constants and $\tilde{\psi}$ is a continuously differentiable function with $\mathcal{H}_{\mathrm{c}}\tilde{\psi}\in L^2([0,1])$, $\tilde\psi( z)=\Or( z^{3/2})$, and $\tilde{\psi}'( z)=\Or( z^{1/2})$.
This leads them to consider self-adjoint realisations of $\mathcal{H}_{\mathrm{c}}$ on subspaces 
\begin{equation}
\mathcal{D}_{\beta_1,\beta_2}=\left\{\psi\in \mathcal{D}_{\mathrm{max}}(\mathcal{H}_{\mathrm{c}})\Big| {\cos\beta_1c_1(\psi)+\sin\beta_1c_2(\psi)=0,\atop \cos\beta_2\psi(1)+\sin\beta_2\psi(1)=0}\right\},
\end{equation}
which are maximal in the sense that
\begin{eqnarray*}
\big\{\psi\in \mathcal{D}_{\mathrm{max}}(\mathcal{H}_{\mathrm{c}})|\langle\mathcal{H}_{\mathrm{c}}\phi|\psi\rangle=\langle\phi|\mathcal{H}_{\mathrm{c}}\psi\rangle\mbox{ for all }\phi\in \mathcal{D}_{\beta_1,\beta_2}\big\}
=\mathcal{D}_{\beta_1,\beta_2}.
\end{eqnarray*}
According to KLP, given any $\beta_2\in[0,\pi)$, one can choose anyone of these maximal subspaces $\mathcal{D}_{\beta_1,\beta_2}$ as domain of $\mathcal{H}_{\mathrm{c}}$ and obtain thereby a self-adjoint realization, where the choice $\beta_1=\pi/2$ corresponds to what is known as Friedrichs realization.

For the special case of the Friedrichs realization, {i.e.\ $\beta_1=\pi/2$} and hence $c_2(\psi)=0$, KLP found that the trace of the resolvent $\Tr(E-\mathcal{H}_{\mathrm{c}})^{-1}$, the heat kernel $\Tr(\rme^{-t\mathcal{H}_{\mathrm{c}}})$, and the zeta function $\zeta(s,\mathcal{H}_{\mathrm{c}})\equiv \Tr(\mathcal{H}_{\mathrm{c}}^{-s})$ exhibit a ``usual'' type of behaviour of the following kind:  If $|E|\to\infty$ , with $E$ belonging to  any sector (of given opening angle) that does not intersect the positive real axis, then the first two quantities have series expansions in powers of $(-E)^{k/2}$ and $t^{(k-3)/2}$, $k=1,2,\ldots,\infty$, respectively. Furthermore, the zeta function $\zeta(s,\mathcal{H}_{\mathrm{c}})$ can be continued from $\Re\, s>1/2$ to a meromorphic function on the complex plane $\mathbb{C}$ with poles at $s=3/2-k$, $k=1,2,\ldots,\infty$. 
However, for all  other self-adjoint extensions ($\beta_1\in[0,\pi)$ with $\beta_1\ne\pi/2$),  logarithmic anomalies appear in the corresponding expansion of these quantities and  $\zeta(s,\mathcal{H}_{\mathrm{c}})$. Thus the latter zeta function have a logarithmic branch point at $s=0$ for such self-adjoint realisations.

Let us now see how these findings relate to our Schr\"odinger problem with singular potentials on the interval $[0,N]$. We are interested in the case where boundary conditions of the form specified in equation~\eqref{eq:c12bc} with $c_2=0$ hold at $ z=0$ and $ z=N$. Owing to the symmetry of the potential with respect to reflections about the mid-plane $ z=N/2$, the operator $\mathcal{H}_v$ commutes with the corresponding reflection operator $R_{\mathrm{mid}}:\psi( z)\mapsto \psi(N- z)$. We can choose eigenfunctions $\regsol_\nu( z)\equiv\regsol( z,\sk_\nu)$ that are even  with respect to these reflections for odd values $\nu=1,3,\ldots,\infty$, but odd with respect to $R_{\mathrm{mid}}$ for even values $\nu=2,4,\ldots,\infty$, by imposing the boundary conditions
\numparts
\begin{eqnarray}
\regsol( z,\sk_\nu)&\mathop{=}\limits_{ z\to 0{+}}&\sqrt{z}[1+\Or( z)],\label{eq:regsolbc1}\\
\regsol( z,\sk_\nu)&\mathop{=}\limits_{ z\to N{-}}&(-1)^{\nu-1}\sqrt{N-z}\,[1+\Or(N- z)]. \quad\label{eq:regsolbc2}
\end{eqnarray}
\endnumparts
The even eigenfunctions then satisfy Dirichlet boundary conditions at the mid-plane $ z=N/2$, $\psi_\nu(N/2)=0$, $\nu=1,3,5,\ldots$; the odd ones Neumann boundary conditions $\psi'_\nu(N/2)=0$, $\nu=2,4,\ldots$. The Hamiltonian $\mathcal{H}_v$ becomes block-diagonal when represented in this basis. Hence, for $v( z)=v^{(\mathrm{sg})}( z)$ with $v_{-1}\equiv 0$, we are back to the situation with $\beta_1=\pi/2$ investigated by KLP, and we are dealing with the Friedrichs extension of the even and odd `blocks' of $\mathcal{H}_{v}$. This means, in particular,  that in the case of a potential which involves only the leading singular term specified in equation~\eqref{eq:vsg} (i.e., $v_{-1}=0$ and no nonsingular contributions), the trace of the resolvent, the heat kernel, and the zeta function $\zeta(s,\mathcal{H}_v)$ should display what KLP termed the ``usual'' behaviour and hence not involve logarithmic anomalies.

However, it is clear that the addition of singular terms of the form $v_{-1}/ z$ will destroy this usual behaviour and lead to the appearance of logarithmic anomalies. As we know from previous work \cite{DGHHRS12,DGHHRS14,DR14} and can also be seen from the results discussed in Sec.~VII of I, the surface free energy $\fs$ has a leading thermal singularity $\asprop m^2 \ln|m|$ whose proportionality coefficient is $\propto v_{-1}${ , where it should be recalled that $m\propto(T-\tc)/\tc$.} Consequently, logarithmic anomalies do occur when $v_{-1}\ne 0$ in quantities such as the surface free energy and the surface excess energy. Owing to the close relationship of the above-mentioned three quantities investigated by KLP with quantities such as the energy and free energy of our model, similar logarithmic anomalies are to be expected in the former when $v_{-1}\ne0$. Although a generalisation of KLP's work to singular potentials~\eqref{eq:vsg} with $v_{-1}\ne0$ appears possible, no such attempt will be made here since this is beyond the scope of the present paper.

With regard to our subsequent considerations let us simply stress that the self-adjoint realisations of $\mathcal{H}_v$ and $\mathcal{H}_{v^{(\mathrm{sg})}}$ we are concerned with are specified by the {boundary conditions~\eqref{eq:regsolbc1} and \eqref{eq:regsolbc2}} and their analogs for $\regsolo( z,\sko)$, respectively, where we assume that both sets of functions $\{\regsol( z,\sk_\nu)\}_{\nu=1}^\infty$ and $\{\regsolo( z,\sko_\nu)\}_{\nu=1}^\infty$ satisfy standard completeness relations, viz.,
\begin{equation}\label{eq:complrel}
\sum_{\nu=1}^\infty\frac{\regsol( z,\sk_\nu)\,\regsol( z^\prime,\sk_\nu)}{\normreg_\nu}=\delta( z- z^\prime).
\end{equation}

We can now follow the steps taken by Gel'fand and Levitan in their original derivation \cite{GL51} of the GL equations to convince ourselves that the PL equations hold also in our case with singular potentials. The starting point is the observation that the PL representation~\eqref{eq:PLrep} for $\regsol( z,\sk)$ can be inverted to obtain a corresponding one for $\regsolo( z,\sk)$,
\begin{equation}\label{eq:iPLrep}
\regsolo( z,\sk)=\regsol( z,\sk)+\int_0^{ z}\rmd{{ z}'}\,\tilde{P}( z,{ z}')\,\regsol({ z}',\sk)
\end{equation}
with a different kernel $\tilde{P}( z, z^\prime)$. {Here the regular solutions $\regsol(z,k)$ and $\regsolo(z,k)$ for arbitrary $k>0$ satisfy the boundary conditions
\begin{eqnarray}\label{eq:bcregsol}
\regsol(z,k)&\mathop{=}_{z\to 0+}&\sqrt{z}[1+\Or(z)],\nonumber\\
\regsolo(z,k)&\mathop{=}_{z\to 0+}&\sqrt{z}[1+\Or(z)].
\end{eqnarray}
Equation~\eqref{eq:iPLrep}} tells us that $\regsolo( z,\sk)$ depends for given $ z\in(0,1)$ on all $\regsol( z^\prime,\sk)$ with $0< z^\prime< z$. 

Following reference~\cite{GL51}, we can now choose $ z$ and $ z^\prime< z$ in the completeness relation \eqref{eq:complrel}. Then the right-hand side vanishes. Let us introduce the scalar product $\langle X,Y\rangle=\sum_{\nu=1}^\infty\normreg_\nu^{-1} X_\nu Y_\nu$ between infinite-dimensional vectors $X=(X_\nu)$ and $Y=(Y_\nu)$ with real-valued components, using the coefficients $1/\normreg_\nu$ defined in equation~\eqref{eq:alphanudef} as (fixed) metric coefficients. The vanishing of the left-hand side of equation~\eqref{eq:complrel} then means that the vectors  
$\mathbf{\Phi}( z)=(\regsol( z,\sk_\nu))$ and $\mathbf{\Phi}( z^\prime)=(\regsol( z^\prime,\sk_\nu))$ are orthogonal to each other, $\langle \mathbf{\Phi}( z),\mathbf{\Phi}( z^\prime)\rangle=0$. Since we know from equation~\eqref{eq:iPLrep} that $\regsolo( z^\prime,\sk_\nu)$ depends only on $\varphi({ z}'',\sk_\nu)$ with ${ z}''<{ z}'$, the vector $(\regsolo( z^\prime,\sk_\nu))$ must also be orthogonal to $\mathbf{\Phi}(z)$. Hence we have
\begin{equation}
\sum_{\nu=1}^\infty\frac{\regsol( z,\sk_\nu)\,\regsolo( z^\prime,\sk_\nu)}{\normreg_\nu}=0.
\end{equation}

Substitution of the PL representation~\eqref{eq:PLrep} then gives
\begin{equation}\fl
\sum_{\nu=1}^\infty\frac{\regsolo( z,\sk_\nu)\,\regsolo( z^\prime,\sk_\nu)}{\normreg_\nu}%\nonumber\\ & \strut 
+\int_0^{ z}\rmd{ z}''\,\kernP( z,{ z}'')\sum_{\nu=1}^\infty\frac{\regsolo({ z}'',\sk_\nu)\,\regsolo( z^\prime,\sk_\nu)}{\normreg_\nu}=0.
\end{equation}

Both contributions to the left-hand side diverge in the limit $ z^\prime\to z$. We can get rid of this divergence by subtracting the completeness relation~\eqref{eq:complrel} for the set of functions $\{\regsolo( z^\prime,\sko_\nu)\}$. This yields the GL equation~\eqref{eq:GLeq} with the kernel $\kernK$ given by \eqref{eq:kernKfi}.

In order to show that the differential {equation~\eqref{eq:diffeqPLkernel}} for $P( z,s)$ remains valid, we proceed along the lines of Rundell \cite{Run97} in his derivation of his equation~(6.6): We subtract the Schr\"odinger equations  $\mathcal{H}_v\regsol( z)=\varepsilon\,\regsol( z)$ and $\mathcal{H}_{\mathring{v}}\regsolo( z)=\varepsilon\, \regsolo( z)$ from each other {and use} the PL representation~\eqref{eq:PLrep} to express $\regsol''( z)$ and $v( z)\,\regsol( z)$ as
\begin{eqnarray}
\regsol''( z)&=&\frac{\rmd^2}{\rmd  z^2}\regsolo( z)+\frac{\rmd}{\rmd  z}[\kernP( z, z)\,\regsolo( z)]%\nonumber\\&&\strut 
+\kernP^{(1,0)}( z, z)\,\regsolo( z)\nonumber\\&&\strut 
{+\int_0^{ z}\rmd{s}\,\kernP^{(2,0)}( z,s)\,\regsolo(s)}
\end{eqnarray}
and
\begin{equation}\fl
v( z)\,\regsol( z)=\mathring{v}( z)\,\regsolo( z)+[v( z)-\mathring{v}( z)] \,\mathring{\regsol}( z)%\nonumber\\&&\strut 
+\int_0^{ z}\rmd{s}\,v( z)\,\kernP( z,s)\,\mathring{\regsol}(s),
\end{equation}
where $\kernP^{(j,k)}( z,s)=\partial_{ z}^j\partial_s^{k}\kernP( z,s)$.
We then integrate the term of $\varepsilon\,(\regsol-\regsolo)=\mathcal{H}_{v}\regsol-\mathcal{H}_{\mathring{v}}\regsolo$ involving second derivatives twice by parts. We thus arrive at
\begin{eqnarray}\label{eq:s0pluseq}
0&=&\left[2\frac{\rmd \kernP( z, z)}{\rmd  z}-[v( z)-\mathring{v}( z)]\right]\regsolo( z)\nonumber\\ &&\strut +\big[\kernP( z,s)\,\partial_s\regsolo(s)-\kernP^{(0,1)}(z,s)\regsolo(s)\big]_{s=0+}\nonumber\\&&\strut
+\int_0^{ z}\rmd{s}\,\regsolo(s)\big\{\kernP^{(2,0)}( z,s)-\kernP^{(0,2)}( z,s)\nonumber \\
&&\strut-\big[v( z)-\mathring{v}(s)\big]\,\kernP( z,s)\big\}.
\end{eqnarray}
If we set $\varepsilon=\sko_\nu^2$, then $\regsolo( z)$ becomes the corresponding eigenfunction $\regsolo_\nu( z)\equiv \regsolo( z,\sko)$. These eigenfunctions  should be complete for the chosen boundary conditions~\eqref{eq:regsolbc1} and \eqref{eq:regsolbc2}. In order that the term $\kernP( z,s)\partial_s\regsolo_\nu( s)|_{s=0+}$ exists, we must have $\kernP( z,s)=\Or(s^{1/2})$ as $s\to 0$. We therefore require that $\kernP$ satisfies the boundary condition
\begin{equation}\label{eq:kernPbc}
\kernP( z,s)=\sqrt{s}\,[Q( z)+\Or(s)].
\end{equation}
This ensures that the contribution $[\ldots]_{s=0+}$ in the second line of equation~\eqref{eq:s0pluseq} vanishes. Hence, both the coefficient of $\regsolo( z)$ in the first line and the one of $\regsolo(s)$ in the integrand must vanish, so that the condition~\eqref{eq:ufromP} and the differential equation~\eqref{eq:diffeqPLkernel} follow.

Unfortunately, finding an analytic solution to the  GL integral equation~\eqref{eq:GLeq} in the full square $[0,N]\times [0,N]$ is not normally possible except in a few special cases. Even in the semi-infinite case where $\kernK( z, z^\prime)$ can be expressed in terms of the eigenfunctions $\regsolo( z,\sk)$ and the Jost function according to equation~\eqref{eq:KkernelJost}, analytic solutions do not appear to be feasible in general.

If one were able to find analytic solutions for the two choices of potentials $v( z)=v^{(\mathrm{sg})}( z)+u( z)$ and $\mathring{v}( z)=v^{(\mathrm{sg})}( z)$ with $u(z)$ of the form specified in equation~\eqref{eq:vTexp} in the finite-interval or half-space case, then $u( z)$ could be reconstructed via equation~\eqref{eq:ufromP} as
\begin{equation}\label{eq:ufromP2}
u( z)=2\frac{\rmd}{\rmd{ z}}\kernP( z, z).
\end{equation}

Rather than aiming at a full analytic solution, we will content ourselves here with a somewhat less ambitious agenda: using appropriate series expansion ans\"atze, we will determine the PL kernel $\kernP( z,s)$ for small values of its arguments $ z$ and $s$. We first consider the simpler case in which the contribution $v_{-1}/ z$ to $v^{(\mathrm{sg})}( z)$ is absent and the Taylor expansion~\eqref{eq:vTexp}
of $u$ involves only powers of $ z^2$.

\subsection{The case of $v_{-1}=0$ and a nonsingular potential  of the form $u=\tilde{u}(z^2)$}
We set $v_{-1}=0$ in \eqref{eq:vsg}, together with all coefficients $v_{2l-1}=0$ with $l=1,2,\dots,\infty$  in the Taylor expansion~\eqref{eq:vTexp} of $u( z)$, and use the ansatz
\begin{equation}\label{eq:Pseriessimple}
\kernP( z,s)=\sqrt{z s}\sum_{j,k=0}^\infty P_{j,k}\,z^{2j}s^{2k}.
\end{equation}
Inserting both series expansions into equation~\eqref{eq:ufromP} and equating the series -expansion terms of orders $ z^0$, $ z^2$, and $ z^4$ yields the conditions
\numparts
\begin{eqnarray}
v_0&=&2P_{00},\\
v_2&=&6(P_{1,0}+P_{0,1}),\\
v_4&=&10(P_{2,0}+P_{1,1}+P_{0,2}).
\end{eqnarray}
\endnumparts
Likewise, matching the expansion terms of orders $\sqrt{z s}$, $z^2\sqrt{z s}$, and $s^2\sqrt{z s}$ in equation~\eqref{eq:diffeqPLkernel} gives
\numparts
\begin{eqnarray}
4P_{1,0}-4P_{0,1}-P_{0,0}v_0&=&0,\\
16P_{2,0}-4P_{1,1}-P_{1,0}v_0-P_{0,0}v_2&=&0,\\
4P_{1,1}-16P_{0,2}-P_{0,1}v_0&=&0.
\end{eqnarray}

{These six} equations for the six unknowns $P_{j,k}$ with $0\le j+k \le2$ can be solved in a straightforward fashion. The results are
\endnumparts
\numparts
\begin{eqnarray}
P_{0,0}&=&\frac{v_0}{2},\\
P_{1,0}&=&\frac{v_2}{12}+\frac{v_0^2}{16},\\
P_{0,1}&=&\frac{v_2}{12}-\frac{v_0^2}{16},\\
P_{2,0}&=&\frac{v_4}{60}+\frac{v_0^3}{384}+\frac{v_0v_2}{32},\\
P_{1,1}&=&\frac{v_4}{15}-\frac{v_0^3}{192}-\frac{v_0v_2}{48},\\
P_{0,2}&=&\frac{v_4}{60}+\frac{v_0^3}{384}-\frac{v_0v_2}{96}.
\end{eqnarray}
\endnumparts

The corresponding expansion of the kernel $\kernK( z,s)$ can then be determined from the GL equation~\eqref{eq:GLeq}. One obtains
\begin{eqnarray}
\kernK( z,s)=&\sqrt{ z s}\bigg[-\frac{v_0}{2}+\left(\frac{v_0^2}{16}-\frac{v_2}{12}\right)( z^2+s^2)\nonumber\\&\strut +\left(\frac{v_0v_2}{96}-\frac{v_0^3}{384}-\frac{v_4}{60}\right)( z^4+s^4)\nonumber\\ &\strut+\left(\frac{v_0v_2}{24}-\frac{v_0^3}{96}-\frac{v_4}{15}\right) z^2s^2+\ldots\bigg].
\end{eqnarray}

\subsection{The case of $v_{-1}\ne0$ and a nonsingular potential $u(z)$}
We proceed to the general case of a potential whose Laurent expansion {about $z=0$} has the form
\begin{equation}
v(z)=-\frac{1}{4z^2}+\frac{v_{-1}}{z}+\sum_{j=0}^{\infty}{v_j}{z^n}.
\end{equation}
We  allow all  coefficients $v_j$, $j=-1,0,\ldots,\infty$, in this expansion to be nonzero and include a contribution $v_{-1}/ z$ in $\mathring{v}( z)=v^{(\mathrm{sg})}( z)$. In this case, the ansatz~\eqref{eq:Pseriessimple} does no longer work and must be generalised. We use instead one of the form
\begin{equation}\label{eq:Pseriesgen}
\kernP( z,s)=\sum_{j=1}^\infty \kernP_l(s/ z)\, z^j,
\end{equation}
where $0<s< z<1$. We insert this ansatz along with the Taylor series~\eqref{eq:vTexp} into equation~\eqref{eq:diffeqPLkernel}, make the change of variable  $s\to \rho=s/ z$ and Laurent expand in $ z$. Matching the terms of order $ z^{-1}$ and $ z^0$ gives us the differential equations
\begin{equation}\label{eq:dglP1}
(1-\rho^{-2})\left[\kernP_1(\rho)+4\rho^2 P_1''(\rho)\right]=0
\end{equation}
and
\begin{equation}\label{eq:dglP2}\fl
(1-\rho^{-1}) v_{-1} P_1(\rho )+\left(\rho^2-1\right) P_2''(\rho )%&\nonumber\\ \strut
-2 \rho\,  P_2'(\rho)+\frac{1}{4} \left(9-\rho^{-2}\right) P_2(\rho )=0.
\end{equation}

Two linearly independent solutions of equation~\eqref{eq:dglP1} are $\sqrt{\rho}$ and $\sqrt{\rho}\ln\rho$. The second one cannot contribute because it would be incompatible with the boundary condition~\eqref{eq:kernPbc} for $\kernP( z,s)$. The coefficient of the first is fixed by  equation~\eqref{eq:ufromP2}. It follows that
\begin{equation}\label{eq:P1res}
P_1(\rho)=\frac{v_0}{2}\sqrt{\rho}.
\end{equation}

We can now insert this result into equation~\eqref{eq:dglP2}. The solution of this differential equation that is consistent with the boundary condition~\eqref{eq:kernPbc} and equation~\eqref{eq:ufromP2} is given by
\begin{equation}\fl
P_2(\rho) =\frac{\pi}{16}(v_1-4v_0v_{-1})\;{}_2F_1(-1/2,-1/2;1;\rho^2)\,\sqrt{\rho}%\nonumber\\ &\strut
 +\frac{v_0v_{-1}(1+\rho)}{2}\sqrt{\rho},
\end{equation}
where ${}_2F_{1}(\alpha,\beta;\gamma;z)$ is a  hypergeometric function. The homogeneous counterpart of the differential equation~\eqref{eq:dglP2} has a second linearly independent solution  $G_{2,2}^{2,0}\big(\rho^2\big|{7/4,7/4 \atop 1/4,1/4}\big)$ given by a Meijer $G$-function. It does not contribute to the solution because its behaviour
\begin{equation}
G_{2,2}^{2,0}\Big(\rho ^2\Big|{7/4,7/4 \atop1/4,1/4}\Big)\mathop{=}_{\rho\to 0}
-\frac{8}{\pi} \sqrt{\rho } \left[2+\ln\frac{\rho}{4}\right]+\Or\big(\rho^{5/2}\big),
\end{equation}
is incompatible with the boundary condition~\eqref{eq:kernPbc}. 
%The coefficient of the other linearly independent solution is fixed by equation~\eqref{eq:ufromP2}.

Upon substituting the series expansion~\eqref{eq:Pseriesgen}  along with  a similar ansatz
\begin{equation}
\kernK( z,s)=\sum_{j=1}^\infty \kernK_j(s/ z)\, z^j
\end{equation}
into the GL integral equation~\eqref{eq:GLeq}, one can solve for the lowest-order term of $\kernK_1$ in a straightforward fashion. The result
\begin{equation}\label{eq:kernK1}
\kernK_1(\rho)=-\frac{v_0}{2}\sqrt{\rho}
\end{equation}
yields
\begin{equation}\label{eq:kernKlo}
\kernK( z,s)\mathop{=}_{0<s< z\ll1}-\frac{v_0}{2}\sqrt{ z s}+\ldots.
\end{equation}

Hence we have
\begin{equation}\label{eq:v0K}
v_0=-2\frac{\rmd \kernK( z, z)}{\rmd  z}\bigg|_{ z=0}=-2\lim_{ z\to 0} z^{-1}\,\kernK( z, z).
\end{equation}

 \section{Trace formulae, their proofs, and consequences}\label{sec:traceform}
 
In this section we will present and prove a trace formulae for Schr\"odinger problems on  finite intervals and the half-line whose potential involves singular terms of the form specified in Sec.~\ref{sec:SPsingpot} [see equations~\eqref{eq:potform} and \eqref{eq:vTexp}] on approaching the boundary. We start by recalling some background.

Trace formulae (see, e.g., \cite{CS89}) allow one to relate properties of the potential to spectral or scattering data. Perhaps the earliest formula of this kind is the following theorem due to Gel'fand and Levitan \cite{GL53}.
\begin{theorem}[Gel'fand-Levitan trace formula]
Let $\varepsilon_\nu$ be the eigenvalues of the operator $-\partial_{ z}^2+v( z)$ on the interval $[0,1]$ subject to Dirichlet boundary conditions, where $v( z)\in {C}^1[0,1]$ satisfies the condition $\int_0^1\rmd{ z}\,v( z)=0$. Then the  trace formula
\[ \frac{1}{4}[v(0)+v(1)]=\sum_{\nu=1}^\infty \left(\nu^2\pi^2-\varepsilon_\nu\right)\]
holds.
\end{theorem}

In I a trace formula for singular potentials of the kind { encountered} in the exact ${n\to\infty}$ solution of the $O(n)$ $\phi^4$ model on a film was needed to relate the coefficient $v_0$ of the nonsingular part of the potential to its scattering data. Here we derive a more general trace formula from which the one exploited in I follows in a straightforward fashion: 

\begin{theorem}[Finite-interval trace formula] \label{th:finttrf}
Let $\mathcal{H}_{\mathring{v}}$ be the analog of the operator defined by equation~\eqref{eq:Hv} for the potential
\begin{equation}\label{eq:v0def}
\mathring{v}( z)=\cases{%
v^{(\mathrm{sg})}( z)&for $0< z<N/2$,\cr
v^{(\mathrm{sg})}(N- z)&for $N/2< z<N$,%
}
\end{equation}
where $v^{(\mathrm{sg})}(z)$ is the singular potential~\eqref{eq:vsg}.  Let 
$\mathring{\varepsilon}_\nu=\mathring{k}_\nu^2$ and $\regsolo( z,\sko_\nu)$, 
 $\nu=1,2,\ldots,\infty$,  denote the  eigenvalues and eigenfunctions of $\mathcal{H}_{\mathring{v}}$ that satisfy the boundary conditions~\eqref{eq:regsolbc1} and ~\eqref{eq:regsolbc2}. Furthermore, let $\regsol( z,\sk_\nu)$ and $\varepsilon_\nu=\sk_\nu^2$ be the corresponding eigenfunctions and eigenvalues of the singular boundary problem specified by equations \eqref{eq:SE}-\eqref{eq:vsymmetry},  \eqref{eq:regsolbc1} and ~\eqref{eq:regsolbc2}. Then the coefficient $v_0$ satisfies the relation
\begin{equation}\label{eq:v0intermsofu}
v_0=2\sum_{\nu=1}^\infty\left(\frac{1}{\normrego_\nu}-\frac{1}{\normreg_\nu}\right)+\frac{2}{N}\int_0^{N/2}\rmd{ z}\big[v( z)-v^{\mathrm{(sg)}}( z)\big],
\end{equation}
where $\normrego_\nu$ and $\normreg_\nu$ are the squares of the $L^2([0,N])$-norms \eqref{eq:alphanudef} of $\regsolo( z,\sko_\nu)$ and $\regsol( z,\sk_\nu)$, respectively.
\end{theorem}

To prove this theorem we start from equation~\eqref{eq:v0K} and use equation~\eqref{eq:kernKfi} for the kernel  $\kernK( z, z)$ to obtain
\begin{equation}\label{eq:Ksmallzdec}
\kernK( z, z)\mathop{=}_{ z\to 0} z\sum_{\nu=1}^\infty\left[\frac{1}{\normreg_\nu}-\frac{1}{\normrego_\nu}\right]+S( z)+\Or( z^2)
\end{equation}
with
\begin{equation}\label{eq:Sdef}
S( z)=\sum_{\nu=1}^\infty\normreg_\nu^{-1}\Big[\regsolo^2( z,\sk_\nu)-\regsolo^2( z,\sko_\nu)\Big].
\end{equation}

Since the small-$z$ behaviour of $S(z)$ is controlled by contributions with $\nu\gg 1$, we can replace the quantities  $\mathring{k}_\nu$,  $k_\nu$, and $\normreg_\nu$ by their asymptotic large-$\nu$ forms. They are conveniently expressed in terms of the momenta
\begin{equation}\label{eq:anudef}
a_\nu=\pi(\nu-1/2)/N.
\end{equation}
We assert that the above quantities behave asymptotically as
\begin{equation}\label{eq:koasexp}
\sko_\nu=a_\nu+\frac{1}{a_\nu}\left\{\frac{1}{2N^2}+\frac{v_{-1}}{N}\left[\ln (4a_\nu N)+\gamma_{\mathrm{E}}\right]\right\}+\Or(a_\nu^{-3}),
\end{equation}
\begin{eqnarray}\label{eq:kasexp}
\sk_\nu&=&a_\nu+\frac{1}{a_\nu}\Biggl\{\frac{1}{2N^2}+\frac{1}{N}\int_0^{N/2}\rmd{ z}\left[v(z)-v^{(\mathrm{c})}( z)\right]\nonumber\\
&&\strut+\frac{v_{-1}}{N}[\ln (4a_\nu N)+\gamma_{\mathrm{E}}]\Bigg\}+\Or\big(a_\nu^{-3}\big),
\end{eqnarray}
and
\begin{equation}\label{eq:alphaasexp}
\normreg_\nu=\frac{N}{\pi a_\nu}+\Or(a_\nu^{-2}),
\end{equation}
where $\gamma_E$ is the Euler-Mascheroni constant.

To show this, note first that the regular solution $\regsolo( z,\sk)$ that satisfies  the boundary condition~\eqref{eq:regsolbc1} at $ z=0$  {becomes for $\sk=\sko_\nu$} the eigenfunction pertaining to the eigenvalue $\sko_\nu^2$. It is given by
\begin{equation}\fl
\regsolo( z,\sk)=\rme^{\rmi\pi/4}\,\frac{1}{\sqrt{2\sk}}\,M_{-\rmi \frac{ v_{-1}}{2\sk},0}(-2\rmi k  z)%\nonumber \\ 
=\sqrt{ z}\;\rme^{\rmi \sk z}\,{}_1F_1\big[\tfrac{1}{2}(1+\rmi v_{-1}/\sk);1;-2\rmi\sk z\big],
\end{equation}
where $M_{\kappa ,\mu }(z)$ denotes a Whittaker $M$-function \cite{AS72,Olver:2010:NHMF,NIST:DLMF}. When $v_{-1}=0$, the result simplifies to
\begin{equation}
\regsolo( z,\sk)=\sqrt{ z}\,J_0(\sk z),\quad v_{-1}=0,
\end{equation}
where $J_\nu(u)$ is a Bessel function of the first kind.

Second, the eigenvalues $\sko_\nu$ follow from the Neumann and Dirichlet boundary conditions the functions $\regsolo( z,\sko_\nu) $ fulfil at $ z=N/2$ for odd  and even values of $\nu$, respectively.

The boundary conditions at $ z=N/2$ yield for $\sko_\nu$ the equations
\begin{equation}\label{eq:1F1bc}
0=\cases{%
M'_{-\rmi \frac{ v_{-1}}{2\sk},0}(-N\rmi \sko_\nu),&$\nu=$ odd,\cr
M_{-\rmi \frac{ v_{-1}}{2\sk},0}(-N\rmi \sko_\nu)=0,&$\nu=$  even,
}
\end{equation}
which simplify to
\begin{equation}\label{eq:kocond}
0=\cases{J_0(N\sko_\nu/2)-N\sko_\nu\,J_1(N\sko_\nu/2),&$\nu=$ odd,\cr
J_0(N\sko_\nu/2),&$\nu=$ even,%
}
\end{equation}
when $v_{-1}=0$.

We first consider the solution of these equations  in the simpler case $v_{-1}=0$. The asymptotic expansions of the roots of $J_0$ can be found  in equation~(10.21.19) of  \cite{NIST:DLMF}. From them, the result given  in equation~\eqref{eq:koasexp} with $v_{-1}$ set to zero follows at once for even $\nu$.  To derive the corresponding ${v_{-1}=0}$~result for odd $\nu$, one can use the familiar asymptotic expansions of the Bessel functions. Equating the leading contribution $\propto \sko^{1/2}_\nu\cos(N\sko_\nu/2-\pi/4)$ of the asymptotic series for the first line of equation~\eqref{eq:koasexp} to zero yields $\sko_\nu=a_\nu+\Or(1/a_\nu)$ for odd $\nu$. Including the next-to-leading term  {$\propto \sko^{-1/2}_\nu$ of this series}, one can expand in $\sko_\nu-a_\nu$ to obtain for the difference the result $1/(2N^2a_\nu)+\Or(1/a_\nu^3)$ in accordance with equation~\eqref{eq:koasexp}.

To do the analogous computation when $v_{-1}\ne 0$, we need the asymptotic large-$\sk$ behaviours of $\regsolo(N/2,\sk)$ and its derivative $\partial_z\regsolo(N/2,\sk)$. These can be determined by using the asymptotic expansion of the function $M_{-\rmi \kappa ,0 }(-\rmi z)$ for $z\to\infty$ given in equation~(13.5.1) of \cite{AS72} and then expanding the truncated series to first order in $v_{-1}$. The calculations are conveniently performed with the aid of {\sc Mathematica}\begin{footnote}{Wolfram Research, Computer code {\sc Mathematica}, version 10.}
\end{footnote}. One obtains
\begin{eqnarray}
\regsolo( z,\sko)&=&\sqrt{\frac{2}{\pi\sko}}\,\Re\Bigg\{\frac{(2\sko z)^{-\rmi\kappa}\Gamma(1/2)}{\Gamma(1/2-\kappa\rmi)}\exp\bigg[\rmi\sko z-\rmi\frac{\pi}{4}+\frac{\pi\kappa}{2}\nonumber\\
&&\strut +\frac{1}{4}(1+2\rmi\kappa)^2\left(-\frac{\rmi}{2\sko z}-\frac{1+\rmi\kappa}{(2\sko z)^2}\bigg)\right]+\Or\big(\sko^{-3}\big)\Bigg\}\nonumber\\ &&\\
&=&\sqrt{\frac{2}{\pi\sko}}\exp\bigg[\frac{\pi v_{-1}}{4\sko}-\frac{1}{16\sko^2 z^2}+\frac{v_{-1}}{4\sko^2  z}+\frac{\pi^2v_{-1}}{16\sko^4}\bigg]\nonumber\\ &&\times\cos\bigg\{\sko z-\frac{\pi}{4}-\frac{1}{8\sko z}-\frac{v_{-1}}{2\sko}\Big[\gamma_{\mathrm{E}}+\ln(8\sko z)\Big]\bigg\}\nonumber\\&&\strut
+\Or\big(\sko^{-7/2}\big).\label{eq:asWhitt}
\end{eqnarray}
The boundary conditions at $ z=N/2$ imply for $\nu=1,3,\ldots,\infty$ that the $\cos\{\ldots\}_{ z=N/2}$ must vanish, and  for $\nu=2,4,\ldots,\infty$ one finds the condition $[\tanh\{\dots\}+\Or(\sko^{-3})]_{ z=N/2}=0$, where the ellipses in both cases represent the argument of the cosine in equation~\eqref{eq:asWhitt}. From these conditions, the result for $\sko_\nu$ given by equations~\eqref{eq:koasexp} and \eqref{eq:anudef} follows at once. 

To derive equation~\eqref{eq:kasexp}, we can use first-order perturbation theory in $u=v-v^{(\mathrm{c})}$ to determine the energy $\varepsilon_\nu=\sk_\nu^2$. This gives
\begin{equation}\fl
\sk_\nu=\sqrt{\sko_\nu^2+\langle\regsolo_\nu|u|\regsolo_\nu\rangle/\normrego_\nu+\ldots}%\nonumber\\ &=&
=\sko_\nu+\frac{1}{2\sko_\nu}\,\langle\regsolo_\nu|u|\regsolo_\nu\rangle/\normrego_\nu+\Or(\sko_\nu^{-2})
\end{equation}
with
\begin{equation}
\langle\regsolo_\nu|u|\regsolo_\nu\rangle=2\int_0^{N/2}u( z)\,\regsolo^2( z,\sko)\,\rmd{ z}.
\end{equation}
Taking into account the familiar identity $\cos^2x=[1+\cos(2x)]/2$, we conclude from the asymptotic expansion~\eqref{eq:asWhitt} that
\begin{equation}
\regsolo^2( z,\sko_\nu)=\frac{1+\cos{\big[\sko_\nu z+\pi/4+\Or(1/\sko_\nu)\big]}}{\pi\sko_\nu}\Big[1+\Or\big(\sko_\nu^{-1}\big)\Big].
\end{equation}
The integral $\int_0^{N/2}\cos[\sko_\nu z +\Or(1)]u( z)\rmd{ z}$ vanishes as $\sko_\nu\to\infty$ by the Riemann-Lebesgue lemma. Consequently, we have
\begin{equation}
\langle\regsolo_\nu|u|\regsolo_\nu\rangle=\frac{2}{\pi\sko_\nu}\int_0^{N/2}u( z)\,\rmd{z}+\Or\big(\sko_\nu^{-2}\big)
\end{equation}
and likewise $\normrego_\nu=N(\sko_\nu\pi)^{-1}+\Or(\sko_\nu^{-2})$. Equation~\eqref{eq:kasexp} follows from these results in conjunction with equation~\eqref{eq:koasexp}. Finally, equation~\eqref{eq:alphaasexp} also follows because perturbative corrections of order $u$ and higher are smaller by at least one factor of $1/a_\nu$.
 
Having established the above results, we can now exploit them to work out the small-$ z$ behaviour of $S( z)$. An elementary calculation gives
\begin{eqnarray}
S( z)&=\frac{\pi}{N}\sum_{\nu=1}^\infty\left[a_\nu(\sk_\nu-\sko_\nu)\partial_{a_\nu}[\regsolo_\nu( z,a_\nu)]^2+\Or(a_\nu^{-1})\right]+\Or( z^2)\nonumber\\
&= z\lim_{ z\to 0}\bigg[\frac{\pi}{N}\sum_{\nu=1}^\infty\left[1+\Or(a_\nu^{-1})\right] \partial_{a_\nu}[\regsolo( z,a_\nu)/\sqrt{ z}]^2\bigg]\nonumber\\ &\phantom{=}\;\;\times \int_0^{1/2}\rmd{ z^\prime}\,u( z^\prime)+\Or( z^2).
\end{eqnarray}
Since  $[\regsolo( z,a_\nu)/\sqrt{ z}]^2$ is a function $f(a_\nu z)$, the required limit becomes that of a Riemann sum  $\lim_{ z\to0}\pi z\sum_\nu\,f'(a_\nu z)$ yielding the integral $\int_0^\infty\rmd{\sk}\,f'(\sk)$. Upon substituting  $v-v^{(\mathrm{c})}$ for $u$ and exploiting the boundary condition~{ \eqref{eq:regsolbc1}} at $ z=0$, we arrive at
\begin{equation}
S( z)=- z\frac{1}{N}\int_0^{1/2}\rmd{ z^\prime}\left[v( z^\prime)-v^{(\mathrm{c})}( z^\prime)\right]+\Or( z^2),
\end{equation}
whose insertion into equation~\eqref{eq:Ksmallzdec} yields equation~\eqref{eq:v0intermsofu} and hence completes the proof of Theorem~\ref{th:finttrf}.

An immediate consequence of Theorem~\ref{th:finttrf} is that the difference $v_0-\tilde{v}_0$ of the potential coefficients $v_0$ and $\tilde{v}_0$ associated with two potentials $v( z)=v^{(\mathrm{c})}( z)+u( z)$ and $\tilde{v}( z)=v^{(\mathrm{c})}( z)+\tilde{u}( z)$ with the same singular part $v^{(\mathrm{c})}( z)$ can be written as
\begin{equation}\label{eq:v0tildev0}
v_0-\tilde{v}_0=2\sum_{\nu=1}^\infty\left(\frac{1}{\tilde{\normreg}_\nu}-\frac{1}{\normreg_\nu}\right)+\frac{2}{N}\int_0^{N/2}\rmd{ z}\big[u( z)-\tilde{u}( z)\big],
\end{equation}
where $\tilde{\normreg}_\nu$ is the analog of $\normreg_\nu$.

We next turn to the derivation of the trace formula for the half-line case used in I. It is not difficult to see that the following theorem  can be inferred from Theorem~\ref{th:finttrf} by taking the limit $N\to\infty$.
\begin{theorem}[Trace formula for half-line case]\label{thm:traceform}
Let $v( z)$ and $\tilde{v}( z)$  be two potentials on the half-line $(0,\infty)$ that vanish faster than $ z^{-1}$ as $ z\to\infty$ and behave as 
\begin{eqnarray}\label{eq:traceform}
v( z\to 0)&=&v^{(\mathrm{sg})}( z)+v_0+o(1),\nonumber\\
\tilde{v}( z\to 0)&=&v^{(\mathrm{sg})}( z)+\tilde{v}_0+o( 1),
\end{eqnarray}
with the same singular part~\eqref{eq:vsg}. Let $\regsol_\nu( z)$, $\nu=1,\ldots,n_b$, be the regular solutions to the Schr\"odinger equation $\mathcal{H}_{v}\regsol_\nu( z)=\varepsilon_\nu\regsol_\nu( z)$ subject to the boundary conditions
\begin{eqnarray}\label{eq:regsolinfbc}
\regsol_\nu( z)&\mathop{=}\limits_{ z\to 0}&\sqrt{ z}[1+\Or( z)]
\end{eqnarray}
that correspond to bound states (where $n_b$ may be zero). 
 Denote by
\begin{equation}\label{eq:normbs}
 \normreg_\nu=\int_0^\infty\regsol^2_\nu( z)\,\rmd{ z}
 \end{equation}
the squares of the $L^2([0,\infty))$ norms of these (real-valued) functions.
Denote furthermore by
$A(\sk)=\rme^{\sigma(\sk)}$ and $\eta(k)$  the scattering amplitude and the scattering 
phase, respectively, which are defined through the large-$z$  asymptotic behaviour of the regular solution
\begin{equation}\label{eq:asbehregsol}
\regsol( z,\sk)\mathop{=}\limits_{ z\to \infty}\frac{A(\sk)}{\sk}\,\sin[\sk  z+\eta(\sk)]+\Or(1/z),
 \end{equation}
in the {absolutely continuous part $k>0$ of the} spectrum.
Furthermore, let $\tilde{\varphi}_{\tilde{\nu}}$, $\tilde{n}_b$, $\tilde{\normreg}_{\tilde{\nu}}$, and $\tilde{A}(\sk)=\rme^{\tilde{\sigma}(\sk)}$ be the  analogous quantities pertaining to the potential $\tilde{v}( z)$. Then the following relation holds 
 \begin{eqnarray}\label{eq:tfv0}\fl
 v_0-\tilde{v}_0&=\frac{4}{\pi}\int_0^\infty\rmd{ \sk}\,\sk^2\left[\rme^{-2\tilde{\sigma}(\sk)}-\rme^{-2\sigma(\sk)}\right]+\sum_{\mathrm{bound}\atop\mathrm{ states }\;\tilde{\nu}}\frac{2}{\tilde{\normreg}_{\tilde{\nu}}}-\sum_{\mathrm{bound}\atop\mathrm{ states }\;\nu}\frac{2}{\normreg_\nu}.
 \end{eqnarray}
 \end{theorem}

To derive this theorem, note first that the integral on the right-hand side of equation~\eqref{eq:v0intermsofu} is $\Or(1/N)$ and hence vanishes as $N\to\infty$. Owing to the symmetry property~\eqref{eq:vsymmetry}, the eigenfunctions for finite $N$ come in pairs that are even or odd  with respect to reflections about the mid-point $N/2$.  For each even (odd) eigenfunction that becomes a  scattering {state} in the limit $N\to\infty$, there is an odd (even) eigenfunction approaching the same scattering state. For either one of these eigenfunctions of any such pair we can use the fact that the regular solutions can be approximated as
\begin{eqnarray}
\regsol( z,\sk_\nu)&\etwa& \frac{1}{\sk_\nu}\,\rme^{\sigma(\sk_\nu)}\,\sin[\sk_\nu  z+\eta(\sk_\nu)],\nonumber\\
\regsolo( z,\sko_\nu)&\etwa& \frac{1}{\sko_\nu}\,\rme^{\sigma(\sko_\nu)}\,\sin[\sko_\nu  z+\eta(\sko_\nu)],
\end{eqnarray}
in the inner region $1\lesssim  z\lesssim N-1$ of the film. The associated norm parameters behave as
\begin{eqnarray}
\normreg_\nu&=&\frac{N}{2k_\nu^2}\,\rme^{2\sigma(k_\nu)}+\Or(1),\nonumber\\ 
\tilde{\normreg}_\nu&=&\frac{N}{2 {\tilde{k}_\nu}^{2}}\,\rme^{2\tilde{\sigma}(\tilde{k}_\nu)}+\Or(1),
\end{eqnarray}
because the error resulting from the differences in the boundary regions is $\Or(1)$. Using this, one concludes that the resulting sum $\frac{1}{N}\sum_\nu\ldots$  of those states $\nu$  that turn into scattering states becomes an integral $\int_0^\infty(\rmd\sk/\pi)\ldots$, which yields the integral on the right-hand side of equation~\eqref{eq:tfv0}.

Next, consider the remaining states (if there are any), which become bound states when $N\to\infty$.  These also come in pairs of even and odd eigenfunctions that approach the same bound state on $[0,\infty)$. The factor of two is canceled by the corresponding factor of $2$ in the norm parameters. Consequently,  equation~\eqref{eq:tfv0} follows, which completes the proof of Theorem~\ref{eq:traceform}. Hence {we arrive at} the following corollary (exploited in I).
\begin{corollary}\label{cor:inftytracef}
Let $v( z)$ and $\tilde{v}( z)$ be two potentials on the half-line $(0,\infty)$ with the  following properties: 
\begin{enumerate} \item[(i)] The Schr\"odinger equation $\mathcal{H}_v\psi( z)=\varepsilon\,\psi( z)$ subject to the boundary condition $\psi( z\to 0)=\Or(\sqrt{ z})$ and its analog with $v\to\tilde{v}$ have no bound-state solutions.
\item[(ii)] Both potentials have the same singular behaviour at $ z=0$ specified in equation~\eqref{eq:vsg}, with identical coefficients $v_{-1}$ and $\tilde{v}_{-1}$ though possibly different limiting values $v_0$ and $\tilde{v}_0$ of their regular parts.
\end{enumerate} 
Then the following relation holds between the difference of the latter coefficients and  $\sigma(\sk)$, the logarithm of the scattering amplitude:
\begin{equation}\label{eq:vvtildesigma}
v_0-\tilde{v}_0=\frac{4}{\pi}\int_0^\infty\rmd{\sk}\,\sk^2\left[\rme^{-2\tilde{\sigma}(\sk)}-\rme^{-2\sigma(\sk)}\right].
\end{equation}
\end{corollary}

\section{Applications of trace formulae}\label{sec:apptrf}

To illustrate the potential of the above trace formulae and the Corollary~\ref{cor:inftytracef}, we now discuss their application to several illustrative cases. Only potentials are considered whose leading near-boundary singularity $-(4z^2)^{-1}$ for $z\to 0$ agrees with that of the self-consistent potential of the exact $n\to\infty$ solution of the $O(n)$ $\phi^4$ theory on a film discussed in \cite{DGHHRS12}, \cite{DGHHRS14} and I.

\subsection{Half-line case with $v(z)=-(4\sinh^2z)^{-1}$}

We begin by discussing the application of Corollary~\ref{cor:inftytracef} to the Schr\"odinger problem on the half-line with the potential 
\begin{equation}\label{eq:vsinh}
v^{(1)}(z)=-\frac{1}{4\sinh^2z}.
\end{equation}
From its Laurent expansion about $z=0$,
\begin{equation}
 \label{eq:v1Lexp}
v^{(1)}(z)=-\frac{1}{4z^2}+\frac{1}{12}-\frac{z^2}{60}+\Or(z^4){,}
\end{equation}
one finds
\begin{equation}\label{eq:v10}
v^{(1)}_0=\frac{1}{12}.
\end{equation}

In order to apply Corollary~\ref{cor:inftytracef}, we choose the auxiliary reference potential $\tilde{v}(z)$ as
\begin{equation}
 \label{eq:v01}
\tilde{v}^{(1)}(z)=-\frac{1}{4z^2}{,}
\end{equation}
so that the corresponding potential parameter $\tilde{v}^{(1)}_0$ vanishes. The same holds for the corresponding coefficients of the other auxiliary potentials $\tilde{v}^{(j)}$, $j=2,3,4$, we shall consider below, i.e.
\begin{equation}\label{eq:v10tilde}
\tilde{v}^{(j)}_0=0,\;\;j=1,2,3,4.
\end{equation}

The regular solution satisfying the Schr\"odinger equation defined by equations~\eqref{eq:SE}, \eqref{eq:Hv}, and \eqref{eq:vsinh}  subject to the boundary condition~\eqref{eq:regsolbc1} is
\begin{eqnarray}\label{eq:regsol1}
\regsol^{(1)}(z,k)&=&\sqrt{\frac{2}{\pi}}\,\frac{\rme^{\pi k}}{\Gamma(1/2+\rmi k)}\,Q_{-1/2}^{\rmi k}(\coth z)\nonumber \\
&=&\sqrt{\tanh (z)} \cosh ^{-\rmi k}(z) \,_2F_1\left(\tfrac{2 \rmi k+1}{4},\tfrac{2 \rmi k+3}{4};1;\tanh^2z\right),\nonumber\\
\end{eqnarray}
where $Q_\nu^\mu(\zeta)$ is an associated Legendre function of the second kind \cite{AS72,Olver:2010:NHMF,NIST:DLMF}. The result given in the second line of equation~\eqref{eq:regsol1} holds for $z>0$ and $k>0$ because $Q_\nu^\mu(\zeta)$  can be written as
\begin{equation}\fl
Q_\nu^\mu(\zeta)=\rme^{\mu\pi\rmi}\frac{\pi^{1/2}\Gamma(\mu+\nu+1)(\zeta^2-1)^{\mu/2}}{2^{\nu+1}\zeta^{\nu+\mu+1}\Gamma(\nu+3/2)}%\nonumber\\&&\times
\,{}_2F_1\big(\tfrac{\nu+\mu}{2}+1,\tfrac{\nu+\mu+1}{2};\nu+\tfrac{3}{2};1/\zeta^2\big)
\end{equation}
when $\zeta \in(1,\infty)$. 

To determine the asymptotic large-$z$ behaviour of $\regsol^{(1)}(z,k)$, we set $t=1-\zeta=1-\coth z=2\rme^{-2z}+\Or(\rme^{-4z})$ and expand in $t$. We thus arrive at the limiting form
\begin{equation}
\regsol^{(1)}(z,k)\mathop{=}_{z\to\infty}\frac{F^{(1)}(-k)\,\rme^{\rmi kz}-F^{(1)}(k)\,\rme^{-\rmi kz}}{2\rmi k}
\end{equation}
with the Jost function
\begin{equation}
 \label{eq:Fv1}
F^{(1)}(k)=\rme^{\sigma^{(1)}(k)-\rmi\eta^{(1)}(k)}=-\rmi\,\frac{2^{1/2}k\,\Gamma(-\rmi k)}{\pi^{1/2}\,\Gamma(1/2-\rmi k)}.
\end{equation}

From the latter, the associated scattering amplitude $A^{(1)}(k)$ can be computed in a straightforward manner via the familiar relation $A(k)=\sqrt{F(k)\,F(-k)}$ between the Jost function $F(k)$ and the amplitude $A(k)$. The result is
\begin{equation} \label{eq:Av1}
 A^{(1)}(k)=\rme^{\sigma^{(1)}(k)}=[(2k/\pi)\coth(k\pi)]^{1/2}.
\end{equation}

The regular solution for the auxiliary Schr\"odinger problem with the potential~\eqref{eq:v2tilde} problem is given by
\begin{equation}
\label{eq:regsolov1}
\tilde{\regsol}^{(1)}(z,k)=\sqrt{z}\,J_0(kz).
\end{equation}
From its asymptotic behaviour
 \begin{equation}
 \tilde{\regsol}^{(1)}(z,k)\mathop{=}_{z\to\infty}\sqrt{\frac{2}{\pi k}}\,\sin(kz+\pi/4)+\Or(1/z)
 \end{equation}
 we can read off the scattering data
\begin{equation}
\label{eq:Atildev1}
\tilde{A}^{(1)}(z)=\rme^{\tilde{\sigma}(k)}=\sqrt{2k/\pi},\quad\tilde{\eta}^{(1)}(k)=\frac{\pi}{4}.
\end{equation}

Upon inserting this result for $\tilde{\sigma}^{(1)}(k)$ and its analog for $\sigma^{(1)}(k)$ given in equation~\eqref{eq:Av1} along with equation~\eqref{eq:v10tilde} into equation~\eqref{eq:vvtildesigma}, we recover indeed the value of $v_0$ stated in equation~\eqref{eq:v10}:
\begin{equation}
 \label{eq:v10tf}
v^{(1)}_0=\frac{4}{\pi}\int_0^\infty\rmd{k}\,k^2\,\frac{\pi}{2k}[1-\tanh(k\pi)]=\frac{1}{12}.
\end{equation}

\subsection{Half-line case with $v^{(\mathrm{sg})}(z)=-(4z^2)^{-1}+4(\pi^2z)^{-1}$}
As second example, we consider the Schr\"odinger problem on the half-line for the potential $v^{(2)}(z)\equiv v(z;\infty,1)-1$, where $v(z;L,m)$ represents the self-consistent potential of I. Recall that the exact analytical form of this potential is not known. However, its following properties established in I determine it  uniquely{:}
\begin{enumerate}
\item 
The small-$z$ behaviour of the potential $v^{(2)}(z)$ is given by equation~\eqref{eq:singpot} with $v_{-1}=4/\pi^2$.
\item For $z\gg 1$, the potential $v^{(2)}(z)$ vanishes exponentially,
\begin{equation}
v^{(2)}(z)= \exp[-2 z+\Or(\ln z)], \quad z\to\infty.
\end{equation}
\item The Schr{\"o}dinger operator $\mathcal{H}_{v^{(2)}}$ on the half-line $0<z<\infty$, defined by equation~ \eqref{eq:Hv}, does not have bound states. 
\item The scattering amplitude $A^{(2)}(\sk)$ determined by the large-$z$ asymptotic behaviour \eqref{eq:asbehregsol} of the corresponding regular solution $\regsol( z,\sk)$ reads
\begin{equation}
A^{(2)}(\sk) =\sqrt{\sk/\arctan \sk}.
\end{equation}
\end{enumerate}

As auxiliary potential, we here choose
\begin{equation}\label{eq:v2tilde}
\tilde{v}^{(2)}(z)=-\frac{1}{4z^2}+\frac{4}{\pi^2z},
\end{equation}
which fulfils equation~\eqref{eq:v10tilde}. As before, it is understood that the boundary condition~\eqref{eq:bcregsol} is imposed on the respective Schr\"odinger equations~\eqref{eq:SE}.

According to I, the regular solution of the Schr\"odinger problem for the auxiliary potential is given by
\begin{equation}\label{eq:regsoltilde2}
\tilde{\regsol}^{(2)}(z,k)=\frac{\rme^{\rmi\pi/4}}{\sqrt{2\sk}}\,M_{-\rmi\kappa,0}(-2\rmi kz) ,\;\;\kappa=\frac{2}{\pi^2k},
\end{equation}
where $M_{-\rmi\kappa,0}(\zm)$ is a Whittaker-$M$ function \cite{AS72,NIST:DLMF,Olver:2010:NHMF}. It behaves as
\begin{equation}\label{eq:regsol2}
  \tilde{\varphi}^{(2)}(z,k)\mathop{=}_{z\to 0}\sqrt{z}\left[1+\frac{4z}{\pi^2}+\left(\frac{4}{\pi^2}-\frac{k^2}{4}\right)z^2+\Or(z^3)\right]
  \end{equation}
for small $z$. Its asymptotic large-$z$ behaviour is of the form
 \begin{equation}\label{eq:largezregsol}
 \tilde{\regsol}^{(2)}(z,k)\mathop{\aseq}_{z\to\infty}\frac{\tilde{A}^{(2)}(k)}{k}\sin[kz+\tilde{\eta}^{(2)}(k,z)]
\end{equation}
with the amplitude
\begin{equation}\label{eq:tildeA}
\tilde{A}^{(2)}(k)=\exp \sigma^{(2)}(k)=\sqrt{k\left[1+\rme^{4/(\pi k)}\right]/\pi}
\end{equation}
and the logarithmic phase shift
\begin{equation}\label{eq:tildeeta}
\tilde{\eta}^{(2)}(k,z)=\frac{\pi}{4}+\arg \Gamma\left(\frac{1}{2}+\frac{2\rmi}{\pi^2k}\right)-\frac{2}{\pi^2k}\ln(2kz).
\end{equation}

The appearance of the $z$-dependent logarithmic term in $\tilde{\eta}(k,z)$, the analog of the phase shift, is due to the  slow decay $\asprop z^{-1}$ of the potential for $z\to\infty$ and well-known from the case of scattering by a Coulomb potential.

These results can be inserted into equation~\eqref{eq:vvtildesigma} to compute $v_0^{(2)}$.   Since the calculation of the required integral has been explained in I, we just quote the result
 \begin{eqnarray}\label{eq:v20}
 v_0^{(2)}&=&\frac{4}{\pi}\int_0^\infty\rmd{k}\,k^2\bigg[\frac{\pi/k}{1+\exp[4/(\pi k)]}-\frac{\arctan k}{k}\bigg]\nonumber\\
&=&\frac{56\zeta(3)}{\pi^4}-1.
 \end{eqnarray}

\subsection{Half-line case with $v^{(\mathrm{sg})}(z)=-(4z^2)^{-1}-4(\pi^2z)^{-1}$}

We next consider the case of the Schr\"odinger problem on the half-line for the case of the self-consistent potential $v^{(3)}(z)\equiv v(z;\infty,-1)$. Again, the exact analytical form of  this potential is not known, but the following properties known from I determine it unambiguously.
\begin{enumerate}
\item 
The small-$z$ behaviour of the potential $v^{(3)}(z)$ is given by equation~\eqref{eq:singpot} with $v_{-1}=-4/\pi^2$.
\item In the limit $z\to\infty$ the potential $v^{(3)}(z)$  vanishes as
\begin{equation}
v^{(3)}(z)\simeq -\frac{1}{2 z^3}, \quad z\to\infty.
\end{equation}
\item The Schr{\"o}dinger operator $\mathcal{H}_{v^{(3)}}$ on the half-line $0<z<\infty$, defined by equation~\eqref{eq:Hv} with the potential $v^{(3)}(z)$, does not have bound states but  a ``half-bound  state'' \cite{Ma06} $\regsol_0( z)\equiv\regsol( z,\sk=0)$  with zero energy. The regular solution  $\regsol_0( z)$
 approaches unity as  $z\to\infty$,  $\lim_{z\to\infty}\regsol_0( z)= 1$.
\item The scattering amplitude $A^{(3)}(\sk)$ for this Schr\"odinger problem with the potential $v^{(3)}(z)$ is given by
\begin{equation}
A^{(3)}(\sk) =\frac{\sk}{\sqrt{1+\pi |\sk|/2}}.
\end{equation}
\end{enumerate}

As reference potential we here choose
\begin{equation}\label{eq:v3tilde}
\tilde{v}^{(3)}(z)=-\frac{1}{4z^2}-\frac{4}{\pi^2z}.
\end{equation}

Thus,
in the case of the Schr\"odinger problem with the potential $v^{(3)}(z)$, the spectrum of the Hamiltonian $\mathcal{H}_v$ is continuous; no discrete (pure point) spectrum exists. However, the  spectrum of the Hamiltonian $\mathcal{H}_{\tilde{v}^{(3)}}$ can be decomposed as $\spek(\mathcal{H}_{\tilde{v}^{(3)}})=\{\tilde{E}_\nu^{(3)}\}\oplus [0,\infty)$ into a pure point part with eigenvalues
\begin{equation}\label{eq:evtilde3}
\tilde{E}^{(3)}_\nu\equiv -q_\nu^2=-\frac{4}{\pi^4(\nu-1/2)^2},\;\;\nu=1,2,\ldots,\infty,
\end{equation}
and a continuous part $[0,\infty)$ of improper eigenvalues $k^2$ indexed by $k\ge 0$. The eigenfunctions associated with the eigenvalues~\eqref{eq:evtilde3} are
\begin{eqnarray}\label{eq:regsol3}
\tilde{\regsol}^{(3)}_\nu(z)&=&\frac{\pi\sqrt{\nu-1/2}}{2}\,M_{\nu-1/2,0}\bigg(\frac{4 z}{\pi^2(\nu-1/2)}\bigg)\nonumber\\
&=&\sqrt{z}\exp\left[-\frac{4z}{\pi^2(2\nu-1)}\right]\,L_{\nu-1}\left[\frac{8z}{\pi^2(2\nu-1)}\right],
\end{eqnarray}
where $L_{\nu-1}(z)$ are Laguerre polynomials. The regular solutions for $k>0$ and $k=0$ are given by
\begin{equation}\label{eq:regsoltilde3}
\tilde{\regsol}^{(3)}(z,k)=\frac{\rme^{\rmi\pi/4}}{\sqrt{2\sk}}\,M_{\rmi \kappa,0}(-2\rmi kz) ,\;\;\kappa=\frac{2}{\pi^2k}
\end{equation}
and
\begin{equation}
\tilde{\regsol}^{(3)}(z,0)=\sqrt{z}\,J_0\big(4z^{1/2}/\pi\big),
\end{equation}
respectively. Both $\tilde{\regsol}^{(3)}_\nu(z)$ and $\tilde{\regsol}^{(3)}(z,k)$ have been normalised such that the boundary condition~\eqref{eq:regsolbc1} is fulfilled. As shown in I, the large-$z$ behaviour of the regular solutions~\eqref{eq:regsol3} is of the form~\eqref{eq:largezregsol} with the scattering amplitude
\begin{equation}\label{eq:tildeA3}
\tilde{A}^{(3)}(k)=\exp \sigma^{(3)}(k)=\sqrt{k\left[1+\rme^{-4/(\pi k)}\right]/\pi}
\end{equation}
and the logarithmic phase shift
\begin{equation}\label{eq:tildeeta3}
\tilde{\eta}^{(3)}(k,z)=\frac{\pi}{4}+\arg \Gamma\left(\frac{1}{2}-\frac{2\rmi}{\pi^2k}\right)+\frac{2}{\pi^2k}\ln(2kz).
\end{equation}
For the norm parameters $\tilde{\normreg}_\nu$ we obtained in I
\begin{equation}
\tilde{\normreg}_\nu=\frac{\pi^4}{64}\,(2\nu-1)^3,\;\,\nu=1,2,\ldots,\infty.
\end{equation}
The summation of the series and the computation of the integral one encounters when applying the trace formula~\eqref{eq:traceform} have been performed in I. We therefore again just quote the result
\begin{eqnarray}\label{eq:v0mincalc}
v_0^{(3)} &=& \frac{4}{\pi}\int_0^\infty\rmd{\sk}\left[\frac{\pi\sk}{1+\exp[-4/(\pi\sk)]}-1-\frac{\pi\sk}{2}\right] %\nonumber\\ &&\strut
+\frac{128}{\pi^4}\sum_{\nu=1}^\infty\frac{1}{(2\nu-1)^3}\nonumber\\ 
&=&\frac{56\,\zeta(3)}{\pi^4},
\end{eqnarray}
so that $v_0^{(3)}$ and $v_0^{(2)}+1$ have identical values according to equation~\eqref{eq:v20}.

\subsection{Check of finite-interval trace formula}

To apply and check the finite-interval trace formula~\eqref{th:finttrf}, we now consider the Schr\"odinger problem on the unit interval $(0,1)$ for the potential
\begin{equation}\label{eq:v4}
v^{(4)}(z)=-\frac{\pi^2}{4\sin^2(\pi z)}.
\end{equation}
From its Laurent series about $z=0$,
\begin{equation}
v^{(4)}(z)=-\frac{1}{4z^2}-\frac{\pi^2}{12}-\frac{\pi^2z^2}{60}+\Or(z^4),
\end{equation}
we read off that
\begin{equation}\label{eq:v40}
v_0^{(4)}=-\frac{\pi^2}{12}.
\end{equation}
It belongs to the class called ``trigonometric Rosen-Morse potentials'' in the literature \cite{CK06,RM32}. 

The solution of our Schr\"odinger problem for this potential can be obtained with the aid  of the solutions of the Schr\"odinger equation for the potential $-U_0/\cosh^2(\alpha x)$ given in \S~23 of  \cite{LL58}. Setting $\alpha=\rmi \pi$, $x=z-1/2$, and $U_0\to \pi^2/4$, imposing the boundary conditions, and taking into account that the eigenfunctions are again either even or odd  with respect to the mid-point $z=1/2$, one can determine the  eigenenergies $E_\nu$ and eigenfunctions  $\varphi_\nu(z)$ in a straightforward fashion. One obtains
\begin{equation}
E_\nu=[\pi(\nu-1/2)]^2,\;\;\nu=1,2,\ldots,\infty,
\end{equation}
and
\begin{equation}
\varphi_\nu(z)=
\sqrt{\frac{\sin(\pi z)}{\pi}}\,P_{\nu -1}[\cos (\pi  z)],
\end{equation}
where the $P_j(x)$ are Legendre polynomials. 

Recalling the familiar orthonormality property 
\begin{equation}
(j+1/2)\int_{-1}^1\rmd{x}\,P_j(x)\,P_{j'}(x)=\delta_{j,j'},
\end{equation}
one sees that the squared norms of the eigenfunctions are given by 
\begin{equation}
\normreg_\nu=\int_0^1\regsol^2_\nu(z)\,\rmd{z}=\frac{1}{\pi^2(\nu-1/2)}.
\end{equation}

As auxiliary potential we choose the potential $\mathring{v}(z)$ defined in equation~\eqref{eq:v0def} with $v_{-1}=0$  and $N=1$, namely
\begin{equation}\label{eq:v4tilde}
\tilde{v}^{(4)}( z)=\cases{%
-\frac{1}{4z^2}&for $0< z<N/2$,\cr
-\frac{1}{4(1-z)^2}&for $N/2< z<N$.%
}
\end{equation}

The eigenfunctions for this auxiliary Schr\"odinger problem satisfying the boundary conditions~{\eqref{eq:regsolbc1} and \eqref{eq:regsolbc2}} are even and odd under reflections about the mid-point $z=1/2$ when $\nu=1,2,\ldots,\infty$ is odd and even, respectively. They read
\begin{equation}
\regsolo_\nu(z)=
\cases{%
\sqrt{z}\,J_0(k_\nu z) &for  $0<z<\frac{1}{2}$,\cr
(-1)^{\nu-1}\,\sqrt{1-z}\,J_0[k_\nu(1-z)]& for $\frac{1}{2}<z<1$,
}
\end{equation}
where the $k_\nu$ are to be determined from the Neumann or Dirichlet boundary conditions at $z=1/2$, depending on whether $\nu$ is odd or even.

Let us use the standard notation $j_{0,m}$ for the $m$th zero of the Bessel function $J_0(k)$ and denote the $m$th zero of the function $J_0(k/2)-k\,J_1(k/2)$ as $\ell_m$, i.e.,
\begin{equation}
J_0(\ell_m/2)-\ell_m\,J_1(\ell_m/2)=0.
\end{equation}
Then we have
\begin{equation}
k_\nu=
\cases{%
2j_{0,\nu/2},&$\nu=2,4,\ldots,\infty,$\cr
\ell_{(\nu+1)/2},&$\nu=1,3,\ldots,\infty$.
}
\end{equation}

For the squared $L^2([0,1])$ norms of the functions $\regsolo_\nu(z)$ one finds
\begin{equation}
\normrego_\nu=[J_0(k_\nu/2)^2+J_1(k_\nu/2)^2]/4.
\end{equation}

The above results can now be inserted into the trace formula~\ref{th:finttrf} to obtain
\begin{eqnarray}
v^{(4)}_0&=&2\sum_{\nu=1}^\infty\bigg(\frac{1}{\normrego_\nu}-\frac{1}{\normreg_\nu}\bigg)+2\int_0^1\rmd{z}\big[v^{(4)}(z)+\mathring{v}(z)\big]\nonumber\\
&=&2\sum_{\nu=1}^\infty\bigg(\frac{1}{\normrego_\nu}-\frac{1}{\normreg_\nu}\bigg)-1=0.822467033\ldots,
\end{eqnarray}
where we evaluated the series $\sum_{\nu=1}^\infty$ numerically using {\sc Mathematica} \cite{Mathematica10}, verifying that the resulting number agrees to $9$ digits with the exact value given in equation~\eqref{eq:v40}. 

\section{Concluding remarks}\label{sec:conclrem}

In this paper we have extended standard tools of inverse scattering theory such as the Povzner-Levitan representation and the Gel'fand-Levitan equation to one-dimensional Schr\"odinger problems  on the half-line and finite intervals whose potentials exhibit  near-boundary singularities of the form~\eqref{eq:singpot} known from the self-consistent potential of the exact $n\to\infty$ solution of the $O(n)$ $\phi^4$ model. As we have seen, the boundary singularities of the potentials imply modifications of the analogs of these equations for nonsingular potentials. 

We have also derived new trace formulae for Schr\"odinger problems on the half-line and finite intervals for such singular potentials. As is borne out by these trace formulae, the boundary singularities entail important modifications of known trace formulae  for nonsingular potentials (cf.\ \cite{CS89}). { The comparison of our trace formula~\eqref{eq:tfv0} with the standard one stated in equation~\eqref{eq:vl0} reveals the differences. In the latter, the potential coefficient $v_0$ is expressed in terms of the scattering phases $\eta_l(k)$ and eigenenergies $\varepsilon_\nu$.  By  contrast, our trace formula~\eqref{eq:tfv0} relates $v_0$ to the amplitude $\sigma(k)$  and norm parameters $\normreg_\nu$.
}
We checked these trace formula or the implied Corollary~\ref{cor:inftytracef} here by applying them to four illustrative cases, for some of which exact results for the potential parameters $v_0^{(j)}$, $j=1,\ldots,4$, are available. In I we benefited from the trace formulae in that they enabled us to determine the potential parameter $v_0$ of the self-consistent potential.

We expect both the inverse scattering techniques for singular potentials and the  trace formulae described here to have useful applications to other interesting problems.

%\bibliographystyle{unsrt}
%\bibliography{bank}

\end{document}